# When few survive to tell the tale: thymus and gonad as auditioning organs: historical overview

Donald R. Forsdyke

*Department of Biomedical and Molecular Sciences, Queen's University, Kingston, Ontario, Canada K7L3N6*

E-mail address: forsdyke@queensu.ca

Outline:

1. Introduction
2. High attrition in seed organs
3. Generation of haplotypic non-uniformity
4. Intergenerational influences on gamete success
5. Changes in properties other than primary function
6. Thymic auditioning – positive anti-near-self and negative anti-self selection
7. Intracellular self/not-self discrimination – peptide sorting at the protein level
8. Ovarian Auditioning?
9. Testicular Auditioning?
10. Pluripotent Stem Cell Auditioning?
11. Conclusions

ABSTRACT

Unlike other organs, the thymus and gonads generate non-uniform cell populations, many members of which perish, and a few survive. While it is recognized that thymic cells are 'audited' to optimize an organism's immune repertoire, whether gametogenesis could be orchestrated similarly to favour high quality gametes is uncertain. Ideally, such quality would be affirmed at early stages before the commitment of extensive parental resources. A case is here made that, along the lines of a previously proposed lymphocyte quality control mechanism, gamete quality can be registered indirectly through detection of incompatibilities between proteins encoded by the grandparental DNA sequences within the parent from which haploid gametes are meiotically derived. This 'stress test' is achieved in the same way that thymic screening for potential immunological incompatibilities is achieved – by 'promiscuous' expression, under the influence of the AIRE protein, of the products of genes that are not normally specific for that organ. Consistent with this, the *Aire* gene is expressed in both thymus and gonads, and AIRE deficiency impedes function in both organs. While not excluding the subsequent emergence of hybrid incompatibilities due to the intermixing of genomic sequences from parents (rather than grandparents), many observations, such as the number of proteins that are aberrantly expressed during gametogenesis, can be explained on this basis. Indeed, promiscuous expression could have first evolved in gamete-forming cells where incompatible proteins would be manifest as aberrant protein aggregates that cause apoptosis. This mechanism would later have been co-opted by thymic epithelial cells which display peptides from aggregates to remove potentially autoreactive T cells.

## 1. Introduction

Those studying gametogenesis have long reported massive cell death within gonads and have suggested this as evidence for a quality control process ensuring that only the "best" gametes emerge (Roosen-Runge 1977; Rodriquez et al. 1997; Findlay et al. 2015). However, regarding possible mechanisms they have been silent. The only organ where comparable cell death occurs is the thymus and here in recent years great progress has been made on mechanism. In particular, there is "promiscuous" thymic expression at the behest of the AIRE protein of genes that are normally expressed in other organs. Remarkably, the AIRE protein is also expressed in gonads where there is also "promiscuous" expression of gene products. There was speculation that promiscuously expressed proteins had "additional functions in cells other than their currently expected cellular roles" (Hecht 1995). This review explores the possibility of a common mechanistic basis for massive cell death in thymus and gonad, which involves the AIRE protein (Schaller et al. 2008).

2. **High attrition in seed organs**

In general, the embryonic development of organs and organ systems involves the differentiation of distinct, but *relatively uniform*, populations of cells. Any cell death during development is part of normal organogenesis and usually involves small, specific, populations that are programmed to die at distinct stages. Even in adult life, for most organs random cell death with replacement by other cells (by division of neighboring cells or through migration) appears to be part of the normal wear-and-tear of existence, and its scale is relatively minor (Danial and Korsmeyer, 2004). However, for two organs – thymus and gonad – massive cell death (atresia) is the norm. They are 'seed organs' in that the thymus seeds the body with T-lymphocytes (Klein et al., 2014), and the gonad seeds gametes. In both cases, cell division within the organ produces a large, but *non-uniform*, cell population. Then 'winners' emerge, and the rest perish. Although attrition can be high in recent thymus emigrants (Van Hoeven et al., 2017), most thymus cell death is organ-based. Likewise, ovarian cell death is generally organ-based. Testicular cells either undergo local deletion or seed spermatozoa to the environment where most die.

For thymus cells the developmental process has been appropriately referred to as 'auditioning' in that some cells, based on selectively advantageous characters, have been found to survive the thymic 'gauntlet,' and the rest die (Forsdyke, 2015). Can gamete-forming cells also be viewed as 'auditioning'? At her birth, a human female has an 'ovarian reserve' of around one million oocytes, which is itself a considerable decline from the five million at 20 weeks gestation. During her lifetime, oocyte number further declines and, from a wave of some 30 ovarian follicles that develop prior to each ovulation, only one is given the opportunity of contributing to a new life (Findlay et al., 2015). A further, less extreme, opportunity for quality control arises when this one egg emerges victorious over other products of meiotic divisions (two polar bodies; Mira, 1998). Are these discardments – the extreme and the less extreme – of potential gamete-forming cells, decided on a random basis? Alternatively, are one or both discardments made on a selective basis, as occurs with thymic cells? Guided by the growing awareness of the precision of intracellular protein aggregations (Forsdyke, 2015), this paper explores possible biological rationales for such selection, both ovarian and testicular. A preliminary report is published elsewhere (Forsdyke, 2016a).

## 3. Generation of haplotypic non-uniformity

The opportunity for specific selection from among a population of cells arises in thymus and gonads because, in contrast to most other organs, their stem cells ultimately generate T lymphocytes or gametes, each with a unique genotype. Cell variability is the result of programmed diversification. For thymic cells, the diversification is largely limited to the genes that contribute to the receptors (TCRs) through which the cells, and those derived from them, recognize specific peptide-MHC (pMHC) complexes on other cells. Within TCR genes, diversification tends to localize to regions that will focus the peptide components of pMHCs (e.g. complementarity determining region 3; CDR3). However, for gonadal cells, the diversification is much more extensive, involving meiotic recombination and the possibility of biased repair processes (gene conversion). Both processes involve base changes and the movement of DNA sequence segments, so generating unique sequence combinations (haplotypes) that may be helpful or harmful, either for the host organism (thymic diversification), or for its future offspring (gonadal diversification).

## 4. Intergenerational influences on gamete success

A child (c) is the product of two parents (p and m) and four grandparents (gps and gms). Failure of development of c, due to inability of the p and m genomes to work together (hybrid inviability), is often due to one or more genic incompatibilities. Clearly the two pairs of corresponding grandparental genomes, based on the alleles expressed, could not have had this problem, otherwise there would have been no p and m to donate their genomes to c.

Indeed, the hybrid *viability* that p and m usually display is an impressive achievement. In each, of the order of perhaps 20,000 genes from a gp productively coexist with a similar number of genes from a gm. Admittedly, for many of these gene pairs (alleles) the sequences may be identical (homozygosity) but, even then, differences in a variety of other factors (e.g. mono or biallelic expression, transcription rates, action of degradative enzymes) can affect the activities, concentrations and locations of final gene products (e.g. proteins).

Yet it is one thing to have a pair of grandparental genomes working together within a potential parent. It is another to have the meiotic recombinational *selections* from those genomes (occurring in p or m gonads), working together in c. For each gamete those 20,000 genes have now been mixed together to generate a unique combination (haplotype) that is unlikely to have preexisted in the history of the species. Stated simplistically, each diploid parent has discarded 10,000 genes from each of the grandparental chromosomal sets it inherited and has united the remaining 10,000 genes from each set to generate a unique haploid assortment of 20,000 genes. The grandparental sets have navigated the respective parental environments successfully, but some aspect of the novel haplotypes they have now generated in p and m gonads might militate against the later success of gamete-borne DNA in c. Thus, there are two major opportunities for incompatibilities between parental gametes, the first due to these meiotic combinations of grandparental DNA sequences, and the second due to untested combinations of the parental DNA sequences themselves (Figure 1).

The incompatibility problem with grandparental sequences should be less than the problem created when two gametes from disparate parents (p and m) – the gene combinations of which have *not* recently navigated a common environment – unite to form a zygote (c). Yet, whereas this hybrid (c) is the seemingly unavoidable result of an interaction between two gametes of independent origins, the existence of a gonadal process for screening the many potential gamete-forming cells, *prior* to their commitment as functioning gametes, could have been advantageous. Thus, some pre-fertilization screening system might have evolved.

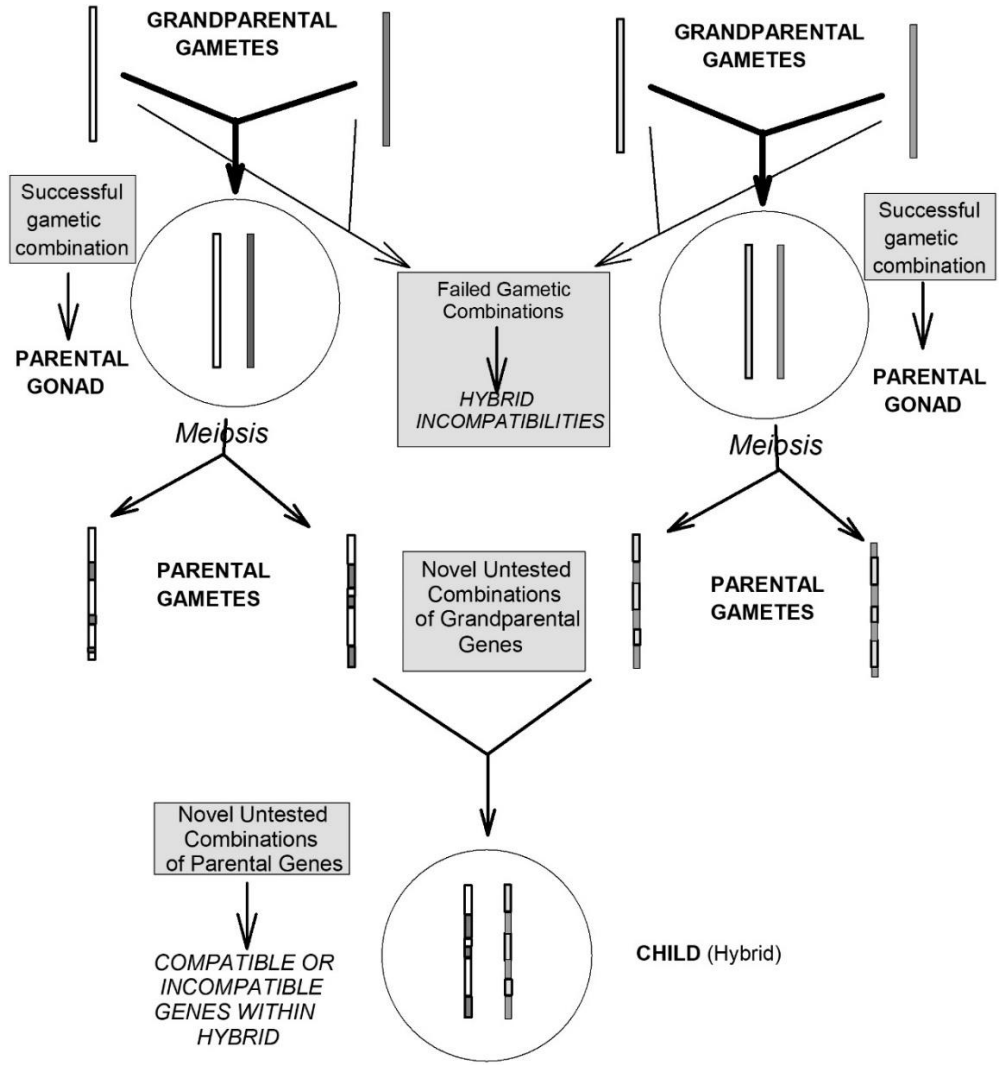

**Fig. 1.** Two opportunities for incompatibilities between parental gametes, the first due to untested meiotic combinations of grandparental DNA sequences, and the second due to untested zygotic combinations of parental DNA sequences. At top, vertical bars with different degrees of uniform shading represent the DNA in grandparental gametes (haploid single chromosomes). Some gametic combinations fail (thin arrows) so that there are no children to grow to become fertile adults (hybrid incompatibilities; i.e. inviabilities). Successful gametic combinations (thick arrows) produce potential parents with gonads where novel combinations of the respective grandparental DNA sequences can be meiotically generated (mixed shading in bars). Both these grandparental combinations, and further combinations arising from a cross between the parents bearing those combinations, may result in hybrid incompatibilities or compatibilities.

This 'audition' would have checked the compatibilities of the unique combinations of grandparental genomes that characterized each parent-borne gamete (the emergent haplotypes). Perhaps at some stage the promiscuous co-expression of diverse genes would have allowed detection of a degree of gene-product incompatibility that otherwise would not normally have been detected in this location (the gonad). Instead the detection would have had to be post-fertilization, perhaps not until an advanced stage of embryo development when the parents had already committed extensive resources to the pregnancy. Thus, a form of early pre-fertilization detection, prior to offspring generation, could have been selectively advantageous, and hence could have evolved.

## 5. Changes in properties other than primary function

The parallels between thymus and gonad that will be drawn here point to another, sometimes overlooked, aspect of gonadal cell diversification. Genetic crosses are often assessed in terms of the extent to which defined functions maintain their continuity between generations. Typically, some externally displayed character is evaluated in terms of its dominance among offspring, or how it is affected by helpful or harmful interactions (epistasis) between paternal or maternal genes. If a genic mutation is not classifiable in terms of a change in a gene's primary function, it may be dismissed as 'neutral,' being not seen as affecting the fitness of offspring.

However, some mutations, while not necessarily affecting the primary function of the protein they encode, can affect its other properties, namely its structure and concentration (Forsdyke 2012a). This could provide an opportunity for an auditioning process, as in the thymus, to sense changes other than malfunction per se. Indeed, while it is considered biologically "unlikely that 'good genes' eggs can be selected, given the low level of gene expression in oocytes" (Mira, 1998), this low level of expression might suffice for the sensing of mutations affecting other properties. These would act as markers, hinting at possible more substantial deleterious changes in that gamete. To assist examination of this, I will first provide a brief outline of thymic auditioning.

## 6. Thymic auditioning – positive anti-near-self and negative anti-self selection

It is generally agreed that there is screening in a 'central' lymphoid organ (bone marrow or thymus) of cells that can give rise to the lymphocytes (B cells and T cells). The majority die and the survivors circulate within the 'peripheral' lymphoid system (blood, lymphatics, lymph-nodes, spleen) from which they may migrate to sites of inflammation. They have been centrally selected based both on their positive responsiveness to 'near-self' antigens (positive selection or "natural autoimmunity;" Forsdyke, 1975, 2012b; Sprent et al. 1988; Cohen, 1992; Cancro and Kearney, 2004), and on their failure to respond strongly to 'self-antigens' (i.e. they escape negative selection; Burnet, 1959). The focus here is on the screening in the thymus of cells that will primarily function in acquired rather than innate immunity (Forsdyke, 2016b).

The thymus was once regarded as having become "vestigial during evolution" and being "just a graveyard for dying lymphocytes" (Miller, 2014). As insightfully predicted by Linsk et al. (1989), it has since been shown that certain antigens that are normally tissue-restricted (TRAs) are synthesized 'promiscuously' in thymic antigen presenting cells (notably epithelial cells; mTECs). The TRAs are then partially degraded to peptides, some of which are displayed as pMHC complexes at mTEC surfaces. Reactions between the TCRs on developing thymus cells and these pMHCs provide sufficient discrimination to eliminate pathological self-reactivity, while providing powerful defence against foreign pathogens (not-self; Takada et al., 2017).

Many TRAs normally operate in non-thymic tissues and there is generally no need for them, although synthesized in the thymus, to express their normal function in that location. Indeed, for certain proteins, their unrestricted functioning within cells in an ectopic organ would damage that organ. It is probably for this reason that mTECs are short-lived – they display peptides from a sample of non-thymic proteins and expire after a few days. But those few days are sufficient for 'self' promiscuous gene products to be *intracellularly* evaluated, not by way of their primary functions, but by way of their components – namely their configurations and, importantly, their concentrations (see later).

In this central location, there are no foreign antigens, so the task at this stage is not to directly discriminate self from not-self, but to discriminate between *three broad categories* of nascent T-cells, (1) those not reacting with self, (2) those reacting with near-self, and (3) those strongly reacting with self. Distinguishing between (1) and (2) in favour of near-self takes advantage of the strategy of foreign organisms (potential pathogens) to mutate closer to self to escape detection (Forsdyke, 1975, 2012b; Cancro and Kearney, 2004). Distinguishing between (2) and (3) in favour of near-self decreases the possibility of autoimmune reactivity (Burnet, 1959; Forsdyke, 2015). Thus, it is now generally agreed that there is "preferential selection and increased survival, in both the thymus and periphery, of T cells that carry specific CDR3 sequences that recognize self-antigens presented by MHC molecules" (Madi et al., 2017; Marrack et al., 2017).

## 7. Intracellular self/not-self discrimination – peptide sorting at the protein level

Certain intracellular structures, proteasomes, specialize in the fragmentation of proteins into short runs of amino acids (peptides), some of which are then bound to MHC complexes (pMHCs). Hedrick (1992) noted: "A troublesome concept concerns the competition among peptides for binding to the available MHC molecules. It is hard to understand how peptides derived from foreign antigens can compete with the tide of self peptides … . Perhaps there is a mechanism that could help to sort peptides into those originating from self and those originating from foreign proteins." The dilemma was reiterated by Gartska et al. (2015): "How a defined MHC I allele selects the correct peptides for presentation out of a large and diverse peptide pool is unclear." Like Hedrick, they supposed that MHC molecules "consider initially a large peptide pool" from which they select. Thus, "Peptides are not chosen at random, but rather are selected for their ability to bind to the polymorphic MHC class I peptide binding groove." However, many self peptides can bind the groove. Indeed, although controversial, much evidence now supports the view that self/not-self discrimination mainly occurs at the protein level, *prior* to the proteosomal processing of proteins into peptides (Colbert et al., 2013; Forsdyke, 2015).

How properties other than its primary function might reveal *intracellularly* whether a protein was derived from self or from not-self, arose from studies of sex-chromosome dosage compensation (Muller, 1948). The need to equalize the dosage of X-borne gene products in females (two X chromosomes) to that of males (one X chromosome), was explained as necessary for maintaining a constant concentration of proteins from generation to generation in the crowded intracellular environment. The purpose of such precision being unclear ("Muller's paradox;" Forsdyke, 1994), it was proposed that exquisite concentration fine-tuning would facilitate the differential aggregation of foreign proteins (or mutated self-proteins), whose concentrations would not be so finely tuned as regular self-proteins (Forsdyke, 1995). The aggregates would be selectively processed by proteasomes into peptides and those peptides for which there was sufficient MHC affinity would then be displayed as pMHC complexes. These complexes, clustered on *peptide-specific* membrane 'rafts,' could then be targeted by peripheral T lymphocytes that had, through their central thymic 'auditioning,' been licensed as less likely to react strongly with self-antigen.

The extensive evidence that has accumulated in support of this hypothesis has been summarized elsewhere (Forsdyke, 2009, 2012c, 2015, 2016a; Vrisekoop et al., 2014; Madi et al., 2017). Indeed, it is recently shown that peptide presentation as a pMHC complex depends more *on the protein it is part of* than on the peptide itself (Pearson et al., 2018). The clustering on rafts of identical pMHC complexes would seem mechanistically easier if, rather than sorting at the pMHC level, there was prior sorting by the formation of specific protein aggregates from which the *colocalized* peptides would then became associated with MHC proteins.

Consistent with the hypothesis was the discovery that the thymic autoimmune regulator (AIRE) protein promotes the 'promiscuous' expression of self-proteins in locations where they are not expected to function. AIRE's binding to chromatin-associated histone-3 proteins is dependent on a position in the histone being unmethylated – an indicator that the associated DNA regions are transcriptionally repressed. Other proteins would then be recruited, including methyl-transferring proteins that would reverse the repression so permitting pervasive ('global') transcription of mRNAs at levels needed for protein synthesis (Abramson et al., 2010).

In this light, fundamental to the thymic auditioning process would be the intracellular synthesis within antigen-presenting mTECs of many self-proteins. Indeed, mTECs are held to "synthesize more proteins than other cell types," possibly leading to "proteotoxic stress" (St-Pierre et al., 2017). Proteins not normally expressed in mTECS should more easily aggregate, so distinguishing them from their fellow-travelers (that were needed for regular thymic functions) and leading to the display of appropriate pMHC complexes. Successful auditioners would be those T cells that reacted weakly, but not strongly, with such pMHC complexes. Strongly-reacting auditioners would die, not because of internal aggregations of their own proteins, but because of prior aggregations within the peptide-presenting cells that they confronted (Forsdyke, 2015).

While for present purposes I deal only with the AIRE protein, it turns out that it is not alone. Thymic promiscuous gene expression of TRAs is not fully abolished when the Aire gene is deleted (Derbinski et al., 2005). Referring to such expression as "beneficial genetic noise," Klein (2015) notes a similar and "Aire-independent and non-redundant role" for Fezf2 (forebrain-expressed zinc finger 2) protein.

## 8. Ovarian auditioning?

Over a female's life time the number of ovarian follicles steeply declines, and only a few emerge 'to tell the tale.' Is the ovary "just a graveyard" for dying eggs? Indeed, Findlay et al. (2015) ponder this as "an evolutionary hangover from aquatic times, when many eggs are shed directly into the environment to ensure that some will be fertilized." But they also consider there might be a mechanism "to ensure that there will be sufficient eggs with a euploid genome, whereas those that have compromised or aneuploid genomes are eliminated."

In humans AIRE deficiency causes autoimmune polyendocrine syndrome type-1 (APS-1) due to failure to eliminate strongly self-reactive T cells (Vogel et al., 2002). Such cells can target diverse organs, including both gonads and the endocrine organs that can influence gonads. Thus, there is a possibility of infertility that is immunologically induced. Indeed, infertility in mutant mice and some APS1 patients was considered to indicate that an "autoimmune syndrome sets in before the reproductive age" (Kyewski and Derbinski, 2004). This implied that the gonadal failure was secondary to thymic failure, rather than being a primary effect. Yet when only thymic *Aire* is impaired, ovarian function remains normal (Dobes et al., 2018).

Since only a few oocytes collectively enter meiosis at one time (Findlay et al., 2015), it might have proven difficult to detect their *Aire* expression. However, histological studies localized *Aire* mRNA expression to growing follicles (Halonen et al., 2001), and Anderson et al. (2002) found that "for tissues other than thymus, lymph node, spleen, and ovary, aire transcripts could not be detected." That gonadal failure in AIRE deficiency might sometimes be primary is also consistent with a report by Nishikawa et al. (2010) that *Aire* mRNA is present in the very early embryo, declining prior to differentiation of the three primary germ layers. They speculated that AIRE played a "role in determining body plan." While acknowledging its presence during spermatogenesis (see later), the possibility that AIRE had been present in the oocyte prior to fertilization was not explored.

As an adjunct to possible screening for DNA repair during this period (Reese and Forsdyke, 2016), perhaps the AIRE-induced 'promiscuous' co-expression of diverse proteins would allow a screening for incompatibilities between them. As with thymocytes, such 'stress tests' might result in cell death (by a process such as apoptosis; Danial and Korsmeyer, 2004), and hence elimination from the competition. Thus, AIRE expression might allow ovarian follicles (and sperm precursors; see later) to globally co-express genes in the segments of parental genomes that had meiotically recombined to generate unique haplotypes. The diverse expression products (proteins) could then be tested against each other to see if coaggregation would occur (since if it occurred here it might also occur later in an offspring). If aggregation occurred the gamete would be deemed to have failed the audition and the individual that might have been created would never come into existence.

How might this work? Figure 2 shows two grandparentally donated chromosomes (A, B) undergoing recombination in the gonad of a prospective parent. The cytoplasmic protein products of two allelic gene pairs (each pair heterologous) are represented by symbols drawn to imply the possibility of negative interactions (deemed not to have occurred, otherwise the prospective parent would not have survived). For simplicity, there is only one recombinational crossover and then the alleles most likely to interact become part of the same haplotype. The mere fact that the alleles are now linked on a chromosome would not seem to have made a difference, since the products appear as capable of cytoplasmic interaction as they were before. However, a cross with another B parent could produce an offspring with an increased concentration of one of the potential reactants. This could suffice to achieve inappropriate aggregation in the hybrid, with ensuing deleterious effects (hybrid inviability).

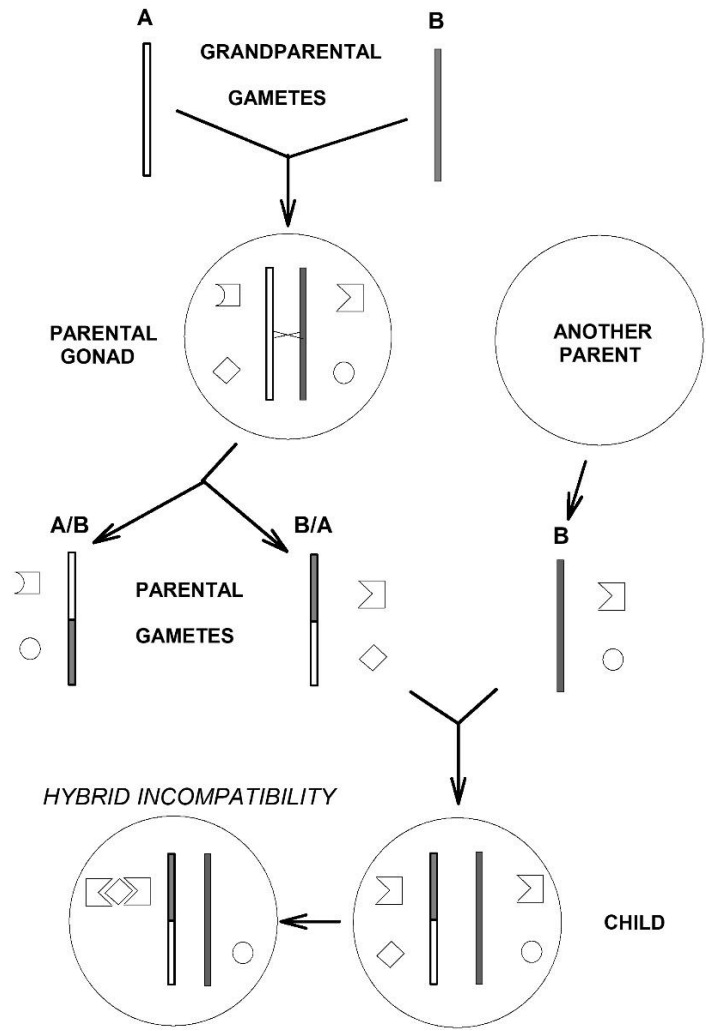

**Fig. 2.** Hybrid incompatibility generated by recombination between grandparental DNA sequences in a parental genome. Vertical bars (A, B) represent haploid single chromosomes within grandparental gametes. Their union creates a diploid prospective parent within whose gonad there is a single recombination event (indicated by a cross connecting the two chromosomes). Symbols within the large circle represent the protein products of two heterozygous allelic gene pairs, with shape suggesting the possibility of cytoplasmic cross-reactivity (epistasis); that with the diamond is stronger and potentially deleterious. Following recombination (generating A/B and B/A), the genes encoding potentially cross-reacting gene products reside on the same chromosomes. Union of one gamete (B/A) with that of another parent carrying a B chromosome, generates a child with one set of homologous allelic genes. This results in a concentration of the corresponding gene product that suffices for its deleterious aggregation with the product of the one member of the other set of allelic genes (still heterologous). Such an outcome would be avoided if during B/A gametogenesis there were promiscuous genic hyperexpression producing the same deleterious aggregation, so B/A gametes would not emerge. For a more complex scenario see Supplementary Materials.

This simple example does little justice to the range and complexity of possible scenarios that might emerge during the 'clash' of genomes when segments of DNA meiotically unite to form a unique gamete (see Supplementary Materials; Appendix 1). Organisms that could have evolved a 'stress test' during parental gametogenesis should have been at a considerable advantage. Under the influence of the AIRE protein, many thousands of proteins would, for a brief period, have come to 'promiscuously' share a common cytoplasm. Irrespective of whether it was itself aggregated, each protein, by virtue of its mere presence in increased quantities, would contribute to the collective pressure to aggregate. Any proteins with a tendency to coaggregate would do so (Forsdyke, 1995). Such aggregation would trigger the death of that particular gamete-forming cell and another, better fitted, gamete would win the day. Given its advantages, stress testing should have evolved early as a fundamental accompaniment to meiosis. Thus, it would be likely to have *preceded* the evolution of the thymus and its associated AIRE-induced promiscuity. The massiveness of gonadal atresias could be explained if 'innocent bystander' cells were drawn into the process (see Supplementary Materials; Appendix 2).

For females, a further opportunity for quality control arises with the extrusion of polar bodies during meiosis (Mira, 1998). It is elsewhere implied that the epigenetic marking of chromosomes, to indicate the number of "suspect" mutations they carry (those having a 50% chance of being deleterious), could provide a basis for differentiating between such meiotic products (Reese and Forsdyke, 2016). Thus, there is likely to be both DNA-level and protein-level input into the gamete auditioning process.

## 9. Testicular auditioning?

Researchers have long been puzzled by gene expression that seems out of place within the testes. When reviewing promiscuous gene expression in the thymus, Kyewski et al. (2002) noted:

> "Expression of tissue-specific genes can also be found during spermatogenesis. The physiology of this observation remains elusive and it has been speculated that it is a by-product of profound epigenetic reprogramming rather than a reflection of a bona fide function of the respective genes in spermatogenesis."

Similar observations in a variety of species had been summarized by Hecht (1995):

"The testis also expresses a large number of proteins that at first glance appear to be surprising candidates for expression during spermatogenesis. … Some researchers explain this apparently promiscuous expression during spermatogenesis as intrinsic to spermatogenesis with widespread transcription of the genome occurring as a result of the changes in chromatin structure needed to produce spermatozoa. … It is equally likely that [the] other proteins … perform additional functions in cells other than their currently expected cellular roles."

Researchers have also been puzzled by the number of potential sperm cells that perish. As with the thymus, the testis was long seen as a graveyard, but Roosen-Runge (1977) observed:

"The loss of germ cells in the course of spermatogenesis is common, regular and often massive, and it raises the question of its *raison d'être.* The phenomenon is apparently closely correlated with the central event of spermatogenesis, meiosis, and with the events which lead up to meiosis, in particular the spermatogonial mitoses. Obviously it is associated in some way with the distribution of genes which regulate phases of the spermatogenic process. Speculations about fundamental reasons for this association are tempting, but not pertinent in the present context. However, the general effect of the cell loss is clearly one of winnowing, of 'quality control,' which serves to select and remove gametes which are in multiple ways not suitable for propagation. By all appearances the effect is large and highly significant, and deserves much more attention than it has received heretofore."

Rodriguez et al. (1997) pondered possible DNA-level "winnowing" mechanisms:

"However, germ cells are extremely sensitive to DNA damage, which is especially incompatible with their ultimate function. It may thus be asked whether the early apoptotic wave may not also be triggered in part by DNA alterations, such as would occur in cells unable to achieve correct DNA rearrangements during the chromosomal crossing over of the pachytene phase of the first meiotic division; this would result in early elimination during development of defective germ cells and of their progenitors arising from a common spermatogonium (because of the unique syncytial nature of this system)."

The latter refers to the sharing by several nuclei of a common cytoplasm (syncytia), where there would be more opportunity for the manifestation of undesirable protein aggregations (Lu et al., 2017). Indeed, even nuclei that have sustained no DNA damage seem to perish in syncytia (Lu and Yamashita, 2017).

As indicated above, Hecht had already mentioned promiscuous expression of proteins within testes. Consistent with this, Schaller et al. (2008) "found Aire protein expressed in the medullary cells of the thymus, in spermatogonia, and in early spermatocytes." They were able to correlate *Aire* gene expression with early waves of apoptosis during spermatogenesis. In their view, such apoptosis, as in the thymus, could be considered as "scheduled" to distinguish it "from 'sporadic' apoptosis events, which can be a consequence of, for example, viral infection, cell stress, or an 'error catastrophe,' i.e., an accumulation of numerous mistakes, in RNAs and/or proteins, severe enough to prevent the cell from functioning." They concluded that "scheduled apoptosis is … a checkpoint of genomic health. Germ cells that fail to complete correct DNA rearrangements during the chromosomal crossing over of the pachytene phase of the first meiotic division, for example, will apoptose." As for promiscuous gene expression, Schaller et al. (2008) acknowledged that:

"It is widely known that in testis one can find expressed sequence tags for numerous genes that have specific functions in other tissues; this promiscuous mRNA expression has been viewed as an inconsequential side effect of the removal of the epigenetic marks on the genome, which is necessary to "reset" the developmental program. Along this line, Aire mRNA might simply be a passenger; in which case its translated product would be without function."

However, it was also acknowledged that such a dramatic process might act as "a counterselection mechanism that keeps the germline stable" by means of some *protein-level* activity:

"We speculate that the promiscuous gene expression found in testis may serve as quality control: Cells with mutated genes would apoptose, perhaps as a consequence of the unfolded protein response. Cells that cannot apoptose … would carry a higher mutational load."

The unfolded protein response has been linked to differential protein aggregation (Sun and Brodsky 2018).

## 10. Pluripotent stem cell auditioning?

Many aspects of embryonic development may be studied using cultured pluripotent embryonic stem cells (ESCs; Evans and Kaufman 1981; Woods and Tilly, 2013; Boroviak and Nichols, 2014). These display constant, seemingly limitless, rapid proliferation while accumulating few abnormalities (i.e. self-renewal rather than differentiation). A major goal has been to find conditions for ESCs to differentiate into specific adult tissues for clinical use, rather than to study the pluripotent state itself.

Since the properties of ESCs closely resemble those of the early ball of multiplying embryonic (epiblast) cells shortly before their uterine implantation, there has been much interest in the discovery that they synthesize AIRE, a property that, as with early embryonic cells (Nishikawa et al., 2010), is lost when they switch to differentiation mode. Likewise, Efroni et al. (2008) noted that the genome of ESCs "is transcriptionally globally hyperactive and undergoes large-scale silencing as cells differentiate". Drawing parallels with the role of AIRE in the thymus, Gu et al. (2010) concluded that AIRE might play an important role "in self-renewal of ESCs and maintaining ESC in a transcriptionally hyperactive state." Thus, they suggested that "mechanisms underlying the promiscuous expression of tissue-specific genes in the immune system may play a similar role in the very diverse gene expression in embryonic stem cells."

However, that the role of AIRE in cultured ESCs might differ from its role in vivo is suggested by recent work showing AIRE to interact with the spindle that controls the separation of chromosomes at cell division (Gu et al., 2017). This is rationalized as follows:

> "As a relatively newly evolved system, the immune system frequently co-opts proteins from basic processes like cell proliferation and embryonic development. … The possibility of a similar, multi-functional role of AIRE is especially intriguing because a central tolerance system coordinated by activating promiscuous gene expression is not required until the evolution of the adaptive immune system and its diversification of immune receptors … . Organisms may have adapted molecules like AIRE from existing biological processes to establish a tolerance mechanism and cope with the random immune receptor diversification."

On the other hand, in keeping with the ability of ESC populations to maintain themselves free from abnormalities, 'promiscuous' global expression from both parental chromosomes (biallelic expression) could create ideal conditions for a further immediate post-fertilization protein-based 'stress test' that might provide another early indication of potential hybrid incompatibilities. This would be consistent with early embryonic expression of *Aire* (Nishikawa et al., 2010), and might be detected as loss of certain cells from self-renewing cultured ESC populations.

## 11. Conclusions

A case can be made that thymus and gonads share the triad of high cell diversity, promiscuous protein expression under the influence of the AIRE protein, and high cell death. Thymus cells that can survive protein-based intracellular self/not-self discrimination 'stress tests' are successful auditioners, optimally poised to face the challenges of foreign or mutated self-antigens. This may be an adaptation of earlier evolving mechanisms by which AIRE facilitates gonadal, and perhaps early embryonic, auditioning to optimally poise offspring to face life's challenges. Some of the experimental approaches to thymus studies might now be profitably applied to gonads. It should be possible to cross disparate mouse types (with or without defects in *Aire* and/or genes mediating cell death) and document the progressive appearance of aggregates in some gamete-forming cells, which then perish. Recent advances in aggregation detection technologies should greatly facilitate this (e.g. Newby et al., 2017).

**Acknowledgements**

The arXiv preprint server (Cornell University) hosts an early copy of this work. Springer Nature gave permission for inclusion of short passages modified from Forsdyke 2016a.

**Supplementary materials**

Two appendices are available below.

# Supplementary Materials

**Appendix 1**  A figure with a more complex hybrid incompatibility scenario than that shown in Figure 2.

**Appendix 2**  Slaughter of innocent bystanders: possible explanation for the generality of the gonadal hyperexpression hypothesis.

**Appendix 1**

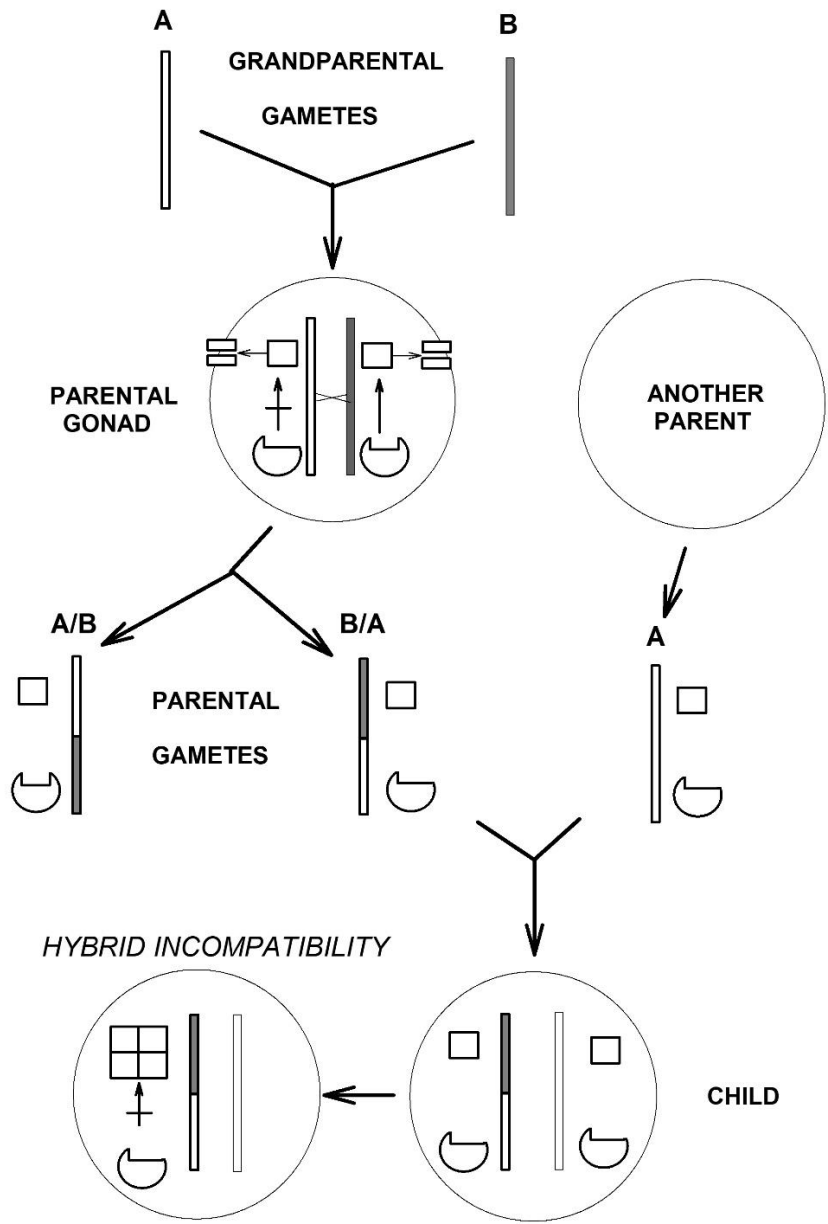

**Fig. 1.** Hybrid incompatibility generated by recombination between grandparental DNA sequences in a parental genome (details as in Fig. 2 in main paper). Here the upper proteins (boxes) are encoded by homozygous alleles in the parent. The lower proteins (circles shaped to imply interaction with upper box proteins) are encoded by heterozygous alleles. The interaction (a specific proteolytic cleavage) initiates the degradation of the box protein, thus lowering its concentration. The circle protein from grandparent A has mutated so that reactivity with box proteins is impaired (indicated by crossed arrow). The circle protein from grandparent B can react with box proteins (indicated by normal arrow), whether they arise from A or B genomes. Since the parent survived, there is haplosufficiency (one circle-encoding allele suffices). Following recombination, A/B gametes are normal, and B/A gametes are abnormal. In the child, box protein degradation is slowed and deleterious aggregates accumulate. Such an outcome would be avoided if during B/A gametogenesis there were promiscuous genic hyperexpression producing the same deleterious aggregation, so that no B/A gametes would emerge as fertilization candidates. During A/B gametogenesis reactivity between box and circle would impede box aggregation even if no proteolytic cleavage occurred. For present purposes neither the diploid genomes of the grandparents and of "another parent," nor the times of mutation in the circle protein, need to be specified.

**Appendix 2**

Slaughter of innocent bystanders: possible explanation for the generality of the gonadal hyperexpression hypothesis.

Tewhey et al. (2011) note the stark "molecular physiological consequences of having variants uniquely distributed across two homologous chromosomal copies of a genomic region" and stress the importance of having more haplotype-resolved genomic sequences (Hoehe et al., 2017). Indeed, it would seem appropriate to refer to a 'clash' of genomes when DNA segments, which may long have been evolutionarily separate, meiotically unite to form a unique gamete. Ideally, deleterious protein interactions that have a *high probability* of occurring in a developing embryo or an adult, would be screened-for during gametogenesis.

However, the *general* hyperexpression of genes in gamete-forming cells, as proposed here, goes further. The proposal would test combinations of proteins that might *not* end up occupying the same organismal compartment. One tissue-specific protein would not normally react intracellularly with a protein specific to another tissue. While the extent of protein tissue-specificity is sometimes challenged (Emig and Albrecht, 2011), nevertheless it would seem undesirable to have stress-tests to detect and eliminate developing gametes that carry the potential for reactivities that are *unlikely to occur*. They are merely innocent bystanders.

It is possible that some elegant mechanism has evolved for activating apoptotic mechanisms *only* when the proteins concerned are *likely* to exist together in the same intracellular environment in a future conceptus. However, the paper makes the case that gonadal auditioning should have evolved *much earlier* than thymic auditioning. In primitive organisms there is less tissue specialization, so a general mechanism could have more easily evolved. Once so evolved, the mechanism could have 'frozen,' so making further refinement impossible. One standard example of this is activation of the larynx in the giraffe that requires a nerve signal to travel down from the brain, loop round the aorta, and then return up the neck to the larynx. Thus, the extensiveness of the culling of gamete-forming cells could be explained if it were of imperfect specificity. Many must die so that few may emerge to tell the tale.